\documentclass[preprint,showpacs,preprintnumbers,amsmath,amssymb,superscriptaddress]{revtex4}
\usepackage{graphicx,graphics}   

\usepackage[utf8]{inputenc}
\usepackage{graphicx,amsfonts}
\usepackage{epstopdf}
\usepackage{subfig}
\usepackage{cancel}
\usepackage{float}
\usepackage[english]{babel}
\usepackage{dcolumn,xcolor,float}
\usepackage{amsmath}
\usepackage{bm}
\begin{document}

\newcommand{\AddrAHEP}{AHEP Group, Institut de F\'{i}sica Corpuscular --
    C.S.I.C./Universitat de Val\`{e}ncia, Parc Cientific de Paterna.\\
    C/Catedratico Jos\'e Beltr\'an, 2 E-46980 Paterna (Val\`{e}ncia) - SPAIN}

\title{ Analytical solution for the Zee mechanism}


\author{A. C. B. Machado}%
\email{a.c.b.machado1@gmail.com}
\affiliation{
Laboratorio de F\'{\i}sica Te\'{o}rica e Computacional\\
Universidade Cruzeiro do Sul -- Rua Galv\~{a}o Bueno 868 \\
S\~{a}o Paulo, SP, Brazil, 01506-000
}

\author{
J. Monta\~no
}%
\email{montano@ift.unesp.br}
\affiliation{
Instituto  de F\'\i sica Te\'orica--Universidade Estadual Paulista \\
R. Dr. Bento Teobaldo Ferraz 271, Barra Funda\\ S\~ao Paulo - SP, 01140-070,
Brazil
}

\author{Pedro Pasquini}
\email{pasquini@ifi.unicamp.br}
\affiliation{~Instituto de F\'isica Gleb Wataghin - UNICAMP, {13083-859}, Campinas SP, Brazil}

\author{V. Pleitez}%
\email{vicente@ift.unesp.br}
\affiliation{
Instituto  de F\'\i sica Te\'orica--Universidade Estadual Paulista (UNESP)\\
R. Dr. Bento Teobaldo Ferraz 271, Barra Funda\\ S\~ao Paulo - SP, 01140-070,
Brazil
}

\date{07/20/2017
%
}
\begin{abstract}
We found an analytical solution for the neutrino mass matrix in the most general case of the Zee model.
Using the recent data on the neutrino parameters besides generating neutrino masses at 1-loop level we fit also the masses of the charged leptons and the leptonic mixing matrix. We also show in what conditions the model is not compatible with neutrino data.
\end{abstract}


\maketitle

\section{Introduction} Among the several ways that neutrinos can gain mass, loop-mechanisms are perhaps the most motivated ones, as they explain naturally the small neutrino masses. Moreover, among all of them, the simplest mechanism to generate neutrino masses at the 1-loop level is the Zee's mechanism~\cite{Zee:1980ai,Cheng:1980qt}. Through the years this mechanism has been studied, see for instance Refs.~\cite{Smirnov:1996bv,Jarlskog:1998uf,Frampton:1999yn,Koide:2001xy,Brahmachari:2001rn,Frampton:2001eu,
Mitsuda:2001vh,Barroso:2005hc,He:2003ih}. 
Here, we will show for the first time an analytical solution for the neutrino mass matrix arising in that mechanism and study the possibility to adjust not only the neutrino masses but also the charged leptons masses and the leptonic mixing matrix using recent data on neutrino masses and the neutrino mixing matrix. 
This mechanism is easy to be implemented in the standard model (SM) by adding an extra scalar doublet and a singly charged singlet scalar, $h^+$.

The representation content of the model in the lepton sector is the usual lepton doublets $\Psi_L=(\nu_a,l_a)^T_L$ and singlets $l_{aR}$, $a=e,\mu,\tau$. The scalar sector consists of two doublets with $Y=+1$, $\phi_{\alpha}=(\phi^+_\alpha\,\phi^0_\alpha)^T$, $\alpha=1,2$ and a singly charged $h^+$ with $Y=+2$. 
As usual $\phi^0_\alpha=(1/\sqrt2)(v_\alpha+\textrm{Re}\phi^0_\alpha+i\textrm{Im}\phi^0_\alpha)$, where $v_\alpha$ are the vaccum spectation values of the neutral fields.

The leptonic Yukawa interactions with the singly charged scalar is
\begin{equation}
-\mathcal{L}^h= \hat{F}_{ab} [\nu^T_{aL} C (V^{l\dagger}_L)_{bk}l^-_{kL} -  (l^-_{L})^T_k(V^{l*}_L)_{ka} C \nu_{bL}]h^+
\label{yukawa1}
\end{equation}
here $C$ is the charge conjugation matrix, $a,b$ are generation indices, and $\hat{F}_{ab}$ is an anti-symmetric matrix, thus $a\not=b$ in (\ref{yukawa1}). Only $h^+$ is still a symmetry eigenstate.

Here, we discuss only the terms that are important for genetation of the neutrino masses. Firstly, notice that there is a trilinear term $f_{\alpha\beta} (\phi_{\alpha}^i \phi_{\beta}^j) \epsilon_{ij} h^-$. The matrix $f_{\alpha\beta}$ is also an anti-symmetric and it has dimension of mass, so we have:
\begin{eqnarray}
-V_h&=&2f_{12}(\phi^+_1\phi^0_2-\phi^+_2\phi^0_1)h^-
\label{trilinear}
\end{eqnarray}
where and $f_{12}$ has dimension of mass. 
We will work in the Higgs basis in which the doublets are written as $\phi_1=[G^+,(1/\sqrt2)(v+\phi^0_1+iG^0)]$ and $\phi_1=[H^+,(1/\sqrt2)(\phi^0_2+iA)]$ and $v=\sqrt{v^2_1+v^2_2}$~\cite{Davidson:2005cw}. In this case the important term Eq.~(\ref{trilinear}) is $-\sqrt{2}f_{12}H^+h^-$. Although, $H^+,h^-$ are not mass eigenstates, we work in this basis because the contribution to neutrino masses are manisfestly finite and the angle which relate the flavor basis $H^-,h^-$ with the mass eigenstates basis $\phi^+_1,\phi^-_2$ is proportional to $f_{12}$. 

Besides the term in Eq.~(\ref{yukawa1}), the more general Yukawa interaction that gives mass to the charged leptons is,
\begin{equation}
-\mathcal{L}^Y_l= 
\sum_{\alpha=1}^2\left[\bar{\nu^\prime}_{aL}(\hat G_\alpha)_{ab} l^{\prime-}_{bR}\phi^+_\alpha+\overline{l^{\prime-}}_{aL}(\hat G_{\alpha})_{ab} l^{\prime-}_{bR}\phi^0_\alpha \right]+H.c.
\label{yukawa2}
\end{equation}
with the charged lepton mass matrix given by $M^l_{ab}=\sum_{\alpha=1}^2 (\hat G^l_{\alpha})_{ab}\frac{v_\alpha}{\sqrt2}$.

There is also an interaction with the physical singly charged scalar, which is related to the lepton masses:
\begin{equation}
-\mathcal{L}=\sum_{\alpha=1}^2\frac{\sqrt2}{v_i}\bar{\nu^\prime}_{a}(\hat G_\alpha)_{ab}^ll^\prime_{bR} \phi^+_2+H.c,
\label{yukaphi+}
\end{equation}
where the primed fields are symmetry eigenstates and neutrinos are massless at the tree level
\section{Solution}
The neutrino are predicted to be majorana particle, whose masses are generated through a loop diagram, which relates them to the lepton masses via the equation~\cite{Wolfenstein:1981kw, Zee:1980ai}, 
\begin{eqnarray}
M^{\nu}&=& \frac{1}{2}\left[\hat{F}M^l\left((M^l)^\dagger-\frac{v_1}{c_\xi}\hat{G}_2^\dagger\right)\right] f_{12} m_h^{-2} \log\left(\frac{m^2_{\phi}}{m^2_{h}}\right)\nonumber \\ &+& {\rm Transpose}
\label{zwh1}
\end{eqnarray}
where $M^l$ is the charged lepton mass matrix in the symmetry basis, $m_h$ is the mass of the charged scalar which is almost singlet. Notice that in the Zee mechanism $\textrm{Tr}M^\nu=0$, hence $m_1+m_2+m_3=0$. We will use this property below. In Eq.~(\ref{zwh1}) we have used the charged lepton mass matrix in order to write the matrix $\hat G_1$ in term of $M^l$ and $\hat G_2$. Notice that in this equation, both $M^\nu$ and $M^l$ are not diagonal. The majorana character of the neutrino mass matrix imply that we can rotate it via unitary rotation of the form $U^T M^\nu U=\hat{M}^\nu\equiv {\rm Diag}[m_1,m_2,m_3]$, while the charged lepton matrix is rotated trough a bi-unitary transformation, $V^l_LM^lV^{l\dagger}_R=m_l\equiv {\rm Diag}[m_e,m_\mu,m_\tau]$. It means that we can re-write Eq.~(\ref{zwh1}) into a simpler form~\cite{He:2003ih},
\begin{eqnarray}
m^\nu&=& \frac{1}{2}F.\left(m^2_l-m_lG_2^\dagger \right) f_{12} m_h^{-2} \log\left(\frac{m^2_{\phi}}{m^2_{h}}\right)\nonumber \\ &+& {\rm Transpose}.
\label{zwh2}
\end{eqnarray}
Now, $F=V^{lT}_L\hat{F}V^l_L$, which is still anti-symmetric and, $G_2^\dagger=\frac{v_1}{c_\xi}V_R\hat{G}^\dagger_2V_L^\dagger$, which is a completly general matrix. Notice that Eq.~(\ref{zwh2}), relates the neutrino mass matrix in the basis where the charged leptons  are diagonal, that is, $U_{\rm PMNS}^Tm^\nu U_{\rm PMNS}=\hat{M}^\nu$, where $U_{\rm PMNS}=(V_L^l)^\dagger U$ is the Pontecorvo–Maki–Nakagawa–Sakata matrix, parametrized by the three neutrino mixing angles, $\sin^2\theta_{12}=0.306\pm0.017$, $\sin^2\theta_{13}0.0214\pm0.0010$, $\sin^2\theta_{23}=0.437\pm0.027 (0.569\pm0.040)$, while the neutrino mass differences, $\Delta m^2_{21}=[7.37\pm0.33]\times10^{-5}$ eV$^2$ and $|\Delta m^2_{31}|=[2.5\pm0.04](2.46\pm0.05)\times10^{-3}$ eV$^2$, can be translated into $\hat M^\nu$ parameters as ${\rm Diag}[(m_1,\pm\sqrt{\Delta m^2_{21}+m^2_1}),\pm\sqrt{\Delta m^2_{31}+m^2_1})]$  for normal or ${\rm Diag}[(\pm\sqrt{\vert\Delta m^2_{13}\vert + m^2_3},\pm\sqrt{ \Delta m^2_{21}+\vert \Delta m^2_{13}\vert+m^2_3},m_3)]$ for inverted hierarchy ~\cite{Olive:2016xmw}.

By writing Eq.~(\ref{zwh2}) in a more suitable form,
\begin{equation} 
m^\nu=(F\hat{Z})+(F\hat{Z})^T,
\label{analitica}
\end{equation}
where $\hat{Z}=\frac{1}{\Lambda}\left(m^2_l-m_lG_2^\dagger \right)$ being a general $3\times3$ matrix with $\Lambda=m_h^2/[f_{12}\log(\frac{m_{\phi}}{m_{h}})]$, one might
think that all the 9 parameters in $\hat{Z}$ contirbute to the neutrino mass matrix $m^\nu$. Nevertheless, it turns out that $\hat{Z}$ can be written as a sum of two matrix $\hat{Z}=Z+Q$ where $Q$ contains all the parameters that do not contribute to Eq.~(\ref{zwh2}) and $Z$ containing only 5 parameters that does contribute to the neutrino mass equation, which we choose to parametrize as,
\begin{equation}\label{eq:Zeq}
Z=\left(
\begin{array}{ccc}
Z_1 & 0 & Z_2\\
0 & Z_3 & 0\\
Z_4 & Z_5 & 0
\end{array}\right)=\left(
\begin{array}{ccc}
-\frac{m^\nu_{13}}{F_{13}} & 0 & -\frac{m^\nu_{33}}{2F_{13}}\\
0 & \frac{F_{12}m^\nu_{33}-2F_{13} m^{\nu}_{23}}{2F_{13}F_{23}} & 0\\
\frac{m^\nu_{11}}{2F_{13}} & \frac{m^\nu_{22}}{2F_{23}} & 0
\end{array}\right)
\end{equation}
In the appendix~\ref{sec:apendiceb}, we describe how to write any general $3\times3$ matrix using this parametrization. Thus, all the parameters that can be used to solve the equation are $5$ from the $Z$ matrix plus $3$ from $F$. That means we can solve the system of equations in Eq.~(\ref{zwh2}) by finding solutions for the five parameters $Z_i$, $i=1,...,5$ and a relation for the three $F_{ij}$ parameters. The analytic solution for $Z_i$ are presented in the second equality of~(\ref{eq:Zeq}) and the relation for $F_{ij}$ can be found in Eq.~(\ref{ultima}).

To show the power of such solution, we will discuss three particular cases where we analyze minimal scenarios of the model: (i) Zee-Wolfenstein Minimal Model, (ii) Next-to-minimal model (iii) Extending the Next-to-minimal.
\begin{itemize}
 \item[(i)] The {\bf  Minimal Zee Model (MZM)} consists on assuming $G_2=0$. Therefore, the only free parameters are the three $F_{ij}$ elements, as the $Z_i$ are fixed by the lepton masses: $Z_1=(m_e^2-m_\tau^2)/\Lambda$, $Z_3=(m_\mu^2-m_\tau^2)/\Lambda$ and the other $Z_i$ equals to zero. 
 
 Moreover, the model predicts the neutrino masses to obey $m_1+m_2+m_3=0$. Unfortunately this model cannot accomodate the neutrino mixing parameters as shown by~\cite{He:2003ih,He:2011hs}, because, in order to fit $\theta_{12}$ and $\theta_{23}$ it cannot produce $\sin^2\theta_{13}<0.33$, which is more than excluded.

\item[(ii)] The {\bf Next-to-Minimal Model} (NtMM) consists on turning on $G_2$ with only one extra parameter $x$. In particular, we found that if we take, $Z_4=-x/F_{13}$, $Z_2=-x/2F_{13}$ and $Z_5=x/2F_{23}$ which preserves the MZM parameters $Z_1=(m_e^2-m_\tau^2)/\Lambda$, $Z_3=(m_\mu^2-m_\tau^2)/\Lambda$ and taking $\delta_{CP}=0$, $m_1+m_2+m_3=0$. Which predicts $x=M_{11}$, plus, taking $\theta_{12}$ and $\theta_{13}$ inside the $3\sigma$ range, it predicts $\tan\theta_{23}=1+{\rm O}(\Delta m_{21}^2/m_2^2)$ for both normal and inverted hierarchy. 
\item[(iii)] A simple {\bf Extension} of the NtMM consists on changing $Z_2\rightarrow-(x-y)/2F_{13}$ and $Z_5\rightarrow(x+y)/2F_{23}$, which gives more freedom to accomodate both $\theta_{23}$ and $\delta_{CP}$. Notice that the NtMM is a particular case for $|y|=0$. The neutrino mass matrix is still traceless, but $m_1+m_2+m_3=0$ does not hold for $\delta_{CP}\neq0$. In Fig.~\ref{fig:resultY} we present the value of the parameter $y$ as a function of $\theta_{23}$. Notice that $|y|=0$ occour around $\theta_{23}=$ maximal, as expected by the result of NtMM.
   \begin{center}
    \begin{figure}[H]
        \centering
        \includegraphics[scale=0.3]{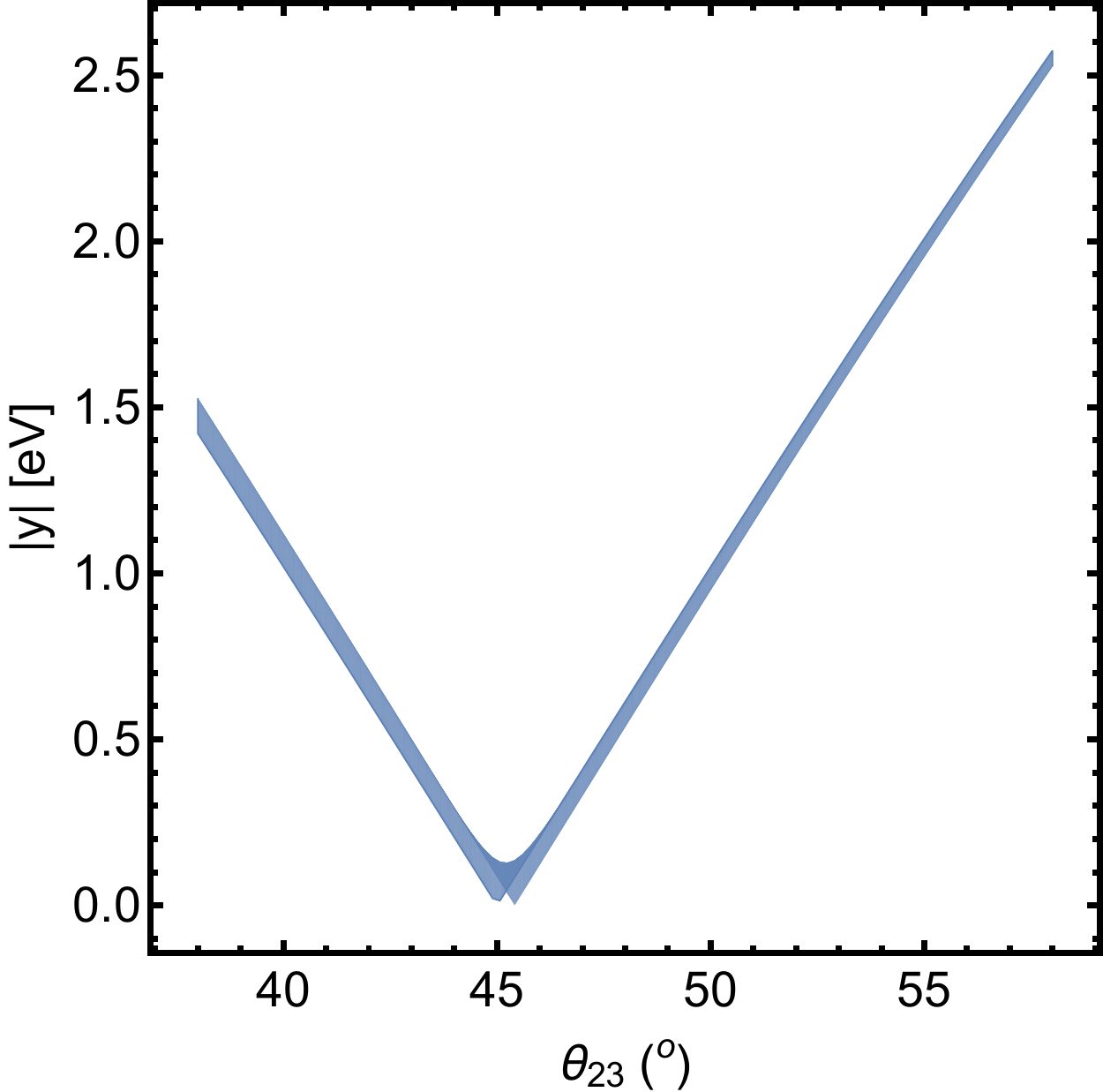}
        \caption{The value of $|y|$ as a function of the atmospheric mixing angle $\theta_{23}$, with central values of $\theta_{12}$ and $\theta_{13}$ given by the global fit and $\delta_{CP}$ from $[0,2\pi]$. The width of the line is due to $\delta_CP$ variation.}
        \label{fig:resultY}
    \end{figure}
\end{center}
\end{itemize}
For both cases (ii) and (iii) the relation~(\ref{ultima}) imply,
\begin{align}
 F_{13}=-\frac{\Lambda}{m_\tau^2-m_e^2} M^\nu_{13}&\\ \nonumber
 \frac{F_{23}(m_\tau^2-m_\mu^2)}{\Lambda}=2M_{13}^\nu& M_{23}^\nu-\frac{F_{12}(m_\tau^2-m_e^2)}{\Lambda}(M^\nu_{11}-y).
\end{align}
The numerical solutions are plotted in Fig.~\ref{fig:resultF} below. Notice that all $F_{ij}$ are of the same order.

\begin{center}
\begin{figure}[H]
\centering
\includegraphics[scale=.6]{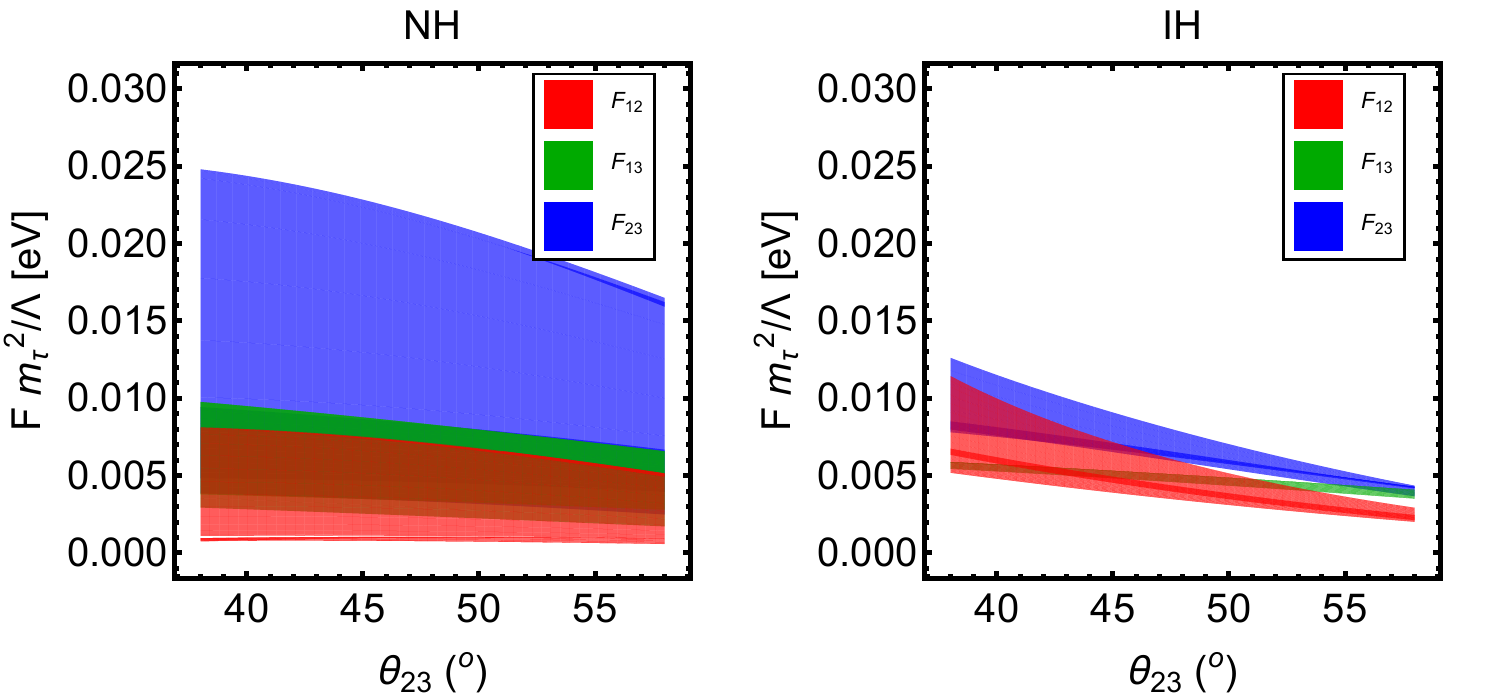}
\caption{The value of $|F_{ij}|$ as a function of the atmospheric mixing angle $\theta_{23}$, with central values of $\theta_{12}$ and $\theta_{13}$ given by the global fit and $\delta_{CP}$ from $[0,2\pi]$. NH(IN) denotes normal(inverted) neutrino mass hierarchy.}
\label{fig:resultF}
\end{figure}
\end{center}

For completeness, we present one possible solution for the mixing matrix $U$, $V_L$ and the $G_2$ Yukawas for case (iii) assuming:
\begin{equation}
 U=U_{\rm PMNS} \qquad V_L=V_R=I_{3\times3}
\end{equation}
and
\begin{equation}
G_2=\Lambda\left(\begin{array}{ccc}
\frac{F_{12}\left(m^\nu_{11}-2y\right) }{4 F_{13} F_{23}m_e} & 0 & -\frac{m^\nu_{11}}{2 F_{13}m_e}\\
0& \frac{F_{12}\left(m^\nu_{11}-2y\right)}{4 F_{13} F_{23}m_\mu}  &  \frac{\left(m^\nu_{11}-2y\right)}{4 F_{23}m_\mu}\\
-\frac{m^\nu_{11}}{2 F_{13}m_\tau} & \frac{\left(m^\nu_{11}-2y\right)}{4 F_{23}m_\tau} & 0
\end{array}\right)
\end{equation}

Notice the special case where $2y=m^\nu_{11}\approx 0.03 (0.05){\rm eV}$ for NH (IH) in which the only non-zero parameters are $(G_2)_{13}=-\frac{m_{11}^\nu\Lambda}{2F_{13}m_e}$ and $(G_2)_{31}=-\frac{m_{11}^\nu\Lambda}{2F_{13}m_\tau}$.

Our result is general, as it does not make any assumption on the values or hierarchy of the coupling constants. For example, take the parameter scan from~\cite{Herrero-Garcia:2017xdu}, there, they assumed $F_{12}=0$ and neglected $m_e$ mass. Nevertheless, the solution we presented gives $(G_2)_{13}/(G_2)_{31}=m_e/m_\tau$, thus, with our result, one may find solutions where the electron mass is important or any parameter is non-zero.

\section{Conclusions}
The Zee model is a simple modification of the standard model Higgs sector with two active Higgs doublets and a singlet singly charged scalar, $h^+$. Here we have considered the most general case when both neutral scalars gain a non-zero vacuum expectation value and the charged lepton mass matrix is not diagonal. Besides explaining the smallness of neutrino masses, it accommodates the charged lepton masses and the leptonic mixing matrix in the $W-l-\nu$ interactions. It also leads to exotic lepton
number violating processes $\vert \Delta L\vert=2$ such as $\bar{\nu}_e+e^-\to \nu_\mu+\mu^-$ via the singly charged scalar.
This induces new interaction of the neutrino with matter that might be constrained by neutrino data. Other processes like $\mu\to e \gamma$, trilepton decays $nn\to pp\mu^-e^-$, $\mu+N\to e+N$,  in general violation of lepton universality processes. One of the neutral Higgs must correspond to the SM Higgs but its trilinear and quartic interactions are modified by loop effects, and the flavor violation Higgs decays as $h\to\mu \tau$,... also occur. Some of these processes have been considered in Ref.~\cite{Herrero-Garcia:2017xdu} but as we noted above these authors did not consider the full space of parameters i.e., their model is not the full Zee model.

\acknowledgments

ACBM thanks  CAPES for financial support.  JM thanks to FAPESP for financial support under the process number  2013/09173-5. VP thanks CNPq for partial support and FAPESP under the funding Grant No. 2014/19164-6. P. P. thanks the support of FAPESP-CAPES funding grant 2014/05133-1,  2014/19164-6 and 2015/16809-9, and the partial support from FAEPEX funding grant,  No 2391/17.

\newpage

\appendix

\section{Appendix}
\label{sec:apendiceb}

The neutrino mass matrix in Eq.~(\ref{analitica}) may be written also as 
\begin{equation}
 m^\nu=F\hat Z+(\hat ZF)^T=FZ+(ZF)^T,
\label{eb1}
\end{equation}
where $\hat Z=Z+Q$,
\begin{equation}\label{eq:Zeq1}
Z=\left(
\begin{array}{ccc}
Z_1 & 0 & Z_2\\
0 & Z_3 & 0\\
Z_4 & Z_5 & 0
\end{array}\right)=\left(
\begin{array}{ccc}
-\frac{m^\nu_{13}}{F_{13}} & 0 & -\frac{m^\nu_{33}}{2F_{13}}\\
0 & \frac{F_{12}m^\nu_{33}-2F_{13} m^{\nu}_{23}}{2F_{13}F_{23}} & 0\\
\frac{m^\nu_{11}}{2F_{13}} & \frac{m^\nu_{22}}{2F_{23}} & 0
\end{array}\right)
\end{equation}
and $Q=\sum_{i=1}^4q_iA_i$, with
\begin{eqnarray}\label{eq:Zeq0}
A_1=\left(
\begin{array}{ccc}
-\frac{F_{23}}{F_{13}}& 0 & 0\\
1 & 0 & 0\\
-\frac{F_{12}}{F_{13}} & 0 & 0
\end{array}\right)\quad
A_2=\left(
\begin{array}{ccc}
0& -\frac{F_{23}}{F_{13}}& 0 \\
0 & 1 & 0\\
0 &-\frac{F_{12}}{F_{13}} & 0 
\end{array}\right)\\ \nonumber
A_3=\left(
\begin{array}{ccc}
0& 0&-\frac{F_{23}}{F_{13}} \\
0 & 0 & 1\\
0& 0 &-\frac{F_{12}}{F_{13}}
\end{array}\right)\quad
A_4=\left(
\begin{array}{ccc}
2& \frac{F_{23}}{F_{13}}&-\frac{2F_{23}}{F_{12}} \\
0 & 1 & \frac{2F_{13}}{F_{12}}\\
0& \frac{F_{12}}{F_{13}} &0
\end{array}\right)
\label{Adef}
\end{eqnarray}
$q_i$ are free parameters. The relation in Eq.~(\ref{eb1}) follows beacuse none of the $A_i$ matrix in (\ref{Adef}) gives non-zero contribution to $m^\nu$.

With  the $Z$ and $A$ matrices given in Eq.~(\ref{eq:Zeq1}) and (\ref{eq:Zeq0}), we found that 
\begin{eqnarray}
&& q_1=\hat{Z}_{21}\quad q_4=\hat{Z}_{33} ,\nonumber \\ &&
q_2=-\frac{2(F_{13})^2 \hat{Z}_{12}-F_{12} F_{23} \hat{Z}_{23}-F_{13} F_{23} \hat{Z}_{33}}{2 F_{13} F_{23} } \nonumber \\ &&
q_3=\frac{F_{12} \hat{Z}_{23}+F_{13} \hat{Z}_{33}}{2F_{13}},\quad 
Z_1=\hat{Z}_{11}-\hat{Z}_{33}+\frac{F_{23} \hat{Z}_{21}-F_{12} \hat{Z}_{23}}{F_{13}} \nonumber \\&&
Z_2=\hat{Z}_{13}+\frac{F_{23}}{F_{13}}\hat{Z}_{23},\quad  
Z_3=\hat{Z}_{22}-\hat{Z}_{33}+\frac{(F_{13})^2\hat{Z}_{12} -F_{12}F_{23} \hat{Z}_{23}}{F_{13}F_{23}} \nonumber \\&&
Z_4=\hat{Z}_{31}+\frac{F_{12}}{F_{13}}\hat{Z}_{21},\quad
Z_5=\hat{Z}_{32}-\frac{F_{12}}{F_{23}}\hat{Z}_{12} 
\end{eqnarray}

Notice that $Z$ has only $5$ parameters, while $m^\nu$ has 6 degrees of freedom. Thus, the missing equations relates $F$,
\begin{eqnarray}
&& F_{23}^2 m^\nu_{11}+2F_{23} F_{12} m^\nu_{13} +F_{13}^2 m^\nu_{22}+F_{12}^2m^\nu_{33}\nonumber \\ &&
-2 F_{13}(F_{23} m^\nu_{12}+F_{12} m^\nu_{23})=0.
\label{ultima} 
\end{eqnarray}

\end{document}